\documentclass[preprint,aps]{revtex4}

\usepackage{graphicx}% Include figure files
\usepackage{dcolumn}% Align table columns on decimal point
\usepackage{bm}% bold math

\begin{document}

\title{Mixing of Pentaquark and Molecular States}

\author{Xiao-Gang He$^{1,2}$, Xue-Qian Li$^1$, Xiang Liu$^1$ and Xiao-Qian
Zeng$^1$}
\affiliation{$^1$Department of Physics, Nankai
University, Tianjin, 300071
\\
$^2$NCTS and Department of Physics, National Taiwan University,
Taipei, 1067 }

\date{\today} %It is always \today, today,

%abstract=================================================================
\begin{abstract}
There are experimental evidences for the existence of narrow
states $\Theta^+$ and $\Theta_c$ with the same quantum numbers of
$uudd\bar s$ and $uudd\bar c$ pentaquarks and also $NK^{(*)}$ and
$ND^{(*)}$ molecular states. Their masses deviate from many
theoretical estimates of the pure pentaquark and molecular states.
In this work we study the possibility that the observed $\Theta^+$
and $\Theta_c$ are mixtures of pure pentaquark and molecular
states. The mixing parameters are in general related to
non-perturbative QCD which are not calculable at present. We
determine them by fitting data from known states and then
generalize the mechanism to $\Theta_b$ to predict its mass and
width. The mixing mechanism can also naturally explain the narrow
width for $\Theta^+$ and $\Theta_c$ through destructive
interferences, even if the pure pentaquark and molecular states
have much larger decay widths. We also briefly discuss the
properties of the partner eigenstates of $\Theta^+$ and $\Theta_c$
and the possibility of experimentally observe them. Moreover,
probable consequences of multi-state mixing are also addressed.

\end{abstract}

\maketitle

\newpage

\section{Introduction}

Following the discovery of $\Theta^{+}$ by the LEPS collaboration
\cite{leps}, some experimental collaborations
\cite{diana,clasa,clasb,saphir,itep,hermes,itep-2,zeus,cosy-tof,togoo,aslanyan,nakano,troyan}
have also confirmed its existence. Its mass is $1539.2\pm 1.6$ MeV
with a very narrow width of $0.9\pm 0.30$ MeV. The $\Theta^{+}$ is
a baryon state with exotic strangeness quantum number $S=+1$ which
cannot be understood as a normal baryon made of three quarks. It
is reasonable to interpret $\Theta^{+}$ as a pentaquark
$(uudd\bar{s})$ which was predicted in several theoretical
works\cite{Predict}. Recently the H1 Collaboration reported their
finding of a new narrow resonance \cite{H1}, whose mass and width
are $3099\pm 3(stat)\pm 5(syst)$ MeV and $12\pm 3$ MeV,
respectively. This narrow resonance can be interpreted as a
charmed pentaquark $\Theta_{c}$ ($uudd\bar c$) which has also been
studied theoretically before\cite{lipkin}. There is also the possibility
of the existence of a new state $\Theta_b$ with the $\bar c$
replaced by a $\bar{b}$ in $\Theta_c$. Even though it has not been
observed at present, future experiments will provide more information.
One should also
note that there are other experiments which do not observe the
$\Theta^+$ and $\Theta_c$\cite{jinshan} states. More
investigations are needed to confirm the existence of these
states.

There have been extensive studies for light pentaquark and
multi-quark states\cite{light,width,Lipkin1,Kingman Cheung}, and
as well as heavy pentquarks\cite{heavy,Lipkin2,Jaffe,Kingman
Cheung}. One of the attractions of investigating pentaquarks is
that one may gain more knowledge on not only the hadron structure,
but also insights to the underlying mechanism which binds quarks
into a multi-body system. It is interesting to investigate if
there exist sub-structures in the five-constituent systems.
Karliner and Lipkin \cite{Lipkin1,Lipkin2} suggested that the
$\Theta^{+}$ has a diquark-triquark $(ud)$-$(ud\bar{s})$
sub-structure, and on the other hand, Jaffe and Wilczek (JW)
\cite{Jaffe} proposed that $\Theta^{+}$ is a bound state of an
antiquark with two highly correlated spin-zero $ud$ diquarks,
moreover they also suggested a mixing of an octet and an
antidecuplet which is recently re-studied \cite{Tetsuo}. In these
frameworks $\Theta^+$ is a $1/2^+$ particle. The predictions on
the central value of $\Theta^+$ mass spread
from\cite{Lipkin1,Lipkin2,Jaffe,Kingman Cheung} 1481 MeV to 1592
NeV and the range covers the central value of the data. The
predictions on the $\Theta_c$ mass is in the range of 2710 MeV to
2997 MeV which is consistently below the central value 3099 MeV
of the data. The mass of $\Theta_b$, using the same method, is
predicted to be in the range of 6050 MeV to 6422 MeV. There are
also several lattice calculations for the masses of the
pentqaurks\cite{lattice,chiu,chiu1} and so far, no conclusive
results about the $\Theta^+$ mass and its parity have been
achieved. For $\Theta_c$ with positive parity the mass is
estimated to be $2977\pm 109$ MeV in Ref. \cite{chiu}. At
present, theoretical estimates have large uncertainties and it is
entirely possible that a pure pentaquark state mass fits the
reported mass of $\Theta_c$ from H1. For the pentaquark decay
width, the situation is even more
uncertain\cite{width,Lipkin1,Lipkin2,Kingman Cheung}. The present
theory is in a very unsatisfactory situation.

There were also attempts to identify $\Theta^+$ as a $N$-$K$
molecular state. However theoretical calculations\cite{
Jaffe,molecular} typically give much larger width and lower mass
for the molecular state compared with the data. There is also a
possibility that the molecular state is a $N$-$K^*$ molecular
state. In this case the mass is above the measured $\Theta^+$
mass, namely a typical negative binding energy of $N$-$K^*$ cannot
reduce the total mass to the data. For this reason, molecular
states cannot be identified as the observed $\Theta^+$. However
these states correspond to a different component in the Hilbert
space, although the triquark-diquark, or diquark-diquark-antiquark
pentaquark combinations and the molecular states have different
color structures, the pentaquark and the moldecular states may mix
because they all have the same overall quantum numbers. It is
clear that no mixing would be needed if the observed states could
be identified with pure pentaquark or molecular states.

There are interesting consequences if mixing indeed exists.
Consider a mixing of two states, a pure pentaquark state mixes
with a molecular state. One notes that when diagonalizing a
two-by-two mass matrix, one obtains two eigenvalues with one of
them being smaller than the minimum of the original two diagonal
matrix elements and another larger than the maximum if the mixing
is non-zero. One of the eigenstates is identified with the
observed $\Theta$ state and another is a physical partner state.
Because mixing, one can expect a mixed state possesses a mass
which is consistent with data, while the predicted pure pentaquark
and the pure molecular $N$-$D$ (or $N$-$D^*$) state have masses
which are different than the observed $\Theta$ state. This
motivates us to consider the possibility that the observed
$\Theta^+$ and $\Theta_c$  may be mixtures of pure pentaquark and
molecular states. Another challenging property of $\Theta^+$ and
$\Theta_c$ is their narrow widths. We will show that even if both
the pure pentaquark and molecular states may have larger widths,
but a destructive interference between them may result in overall
narrow widths for the observed resonances.

Similar idea in obtaining a narrow width for other systems was
discussed in \cite{ref-n} and some authors suggested that the
smallness of the width of $\Theta^+$ may be due to a so-called
``super-radiance'' which actually is also a destructive
interference effect\cite{Auerbach}.

Although at this stage the indication of mixing is not strong,
nevertheless it is interesting to see what this will lead to. In
this work we study some consequences of pure pentaquark and
molecular states mixing for $\Theta^+$, $\Theta_c$ and $\Theta_b$.
Our strategy is as the following. We first calculate the mass of
the $N$-$K$ ($N$-$K^*$)molecular state (having the same quantum
number as $\Theta^+$)  by using linear $\sigma-$model and taking a
theoretical prediction for the pure pentaaquark $\Theta^+$ mass as
input for the mixing mass matrix. We phenomenologically introduce
a mixing parameter in the two-state mass matrix, and diagonalize
the mass matrix to obtain new eigenvalues and eigenstates. By
fitting data, we determine the mixing parameter with which we
evaluate the total width of the corresponding eigenstate.

Indeed, our discussions cannot offer explanations for large mixing
between a pentaquark and a molecular state which requires a good
understanding of non-perturbative QCD effects. We will stay at the
phenomenological level to study the consequences. More accurate
experimental measurements and lattice QCD calculations on
properties of the resonances may provide some clues to this
problem.

We then carry out calculations for $\Theta_c$ with the same
strategy and determine the corresponding mixing parameter by
fitting data. Because charm quark is much heavier than strange
quark, one cannot expect the mixing parameters in the cases for
$\Theta^+$ and $\Theta_c$ to have any direct relation. However,
bottom and charm quarks all are supposed to be heavy compared with
the QCD scale, thus there may be a connection between the
parameters for $\Theta_c$ and $\Theta_b$. By a simple argument
based on one gluon exchange picture we relate the parameters for
$\Theta_b$ to those of $\Theta_c$. Using this value, we estimate
the mass and width for $\Theta_b$.

Obviously there could be multi-state mixing among pentaquark and
molecular states of N-P and N-V types. By adjusting parameters
(there are more of them than in the two-state mixing), the
measured values can be re-produced. If none of the pure states has
a mass closer to the observed pentaquark states, the mixing
parameters need to be large. This is the case we are interested
in. Using model calculations based on one particle ((pseudo)scalar
or vector meson) exchange, we find that mixing between N-P and N-V
states is considerably smaller than the mixing parameter of
pentaquark with either P-N or V-N which is obtained by fitting
data. We therefore will only concentrate on the mixing between the
pentaquark and  molecular states. We will analyze the simple
two-state mixing case in details, and then will discuss the
possible multi-state mixing.
%The corresponding results are
%illustrated in a few figures where we show the changes brought up
%by the three-state mixing.

This paper is organized as follows, after the introduction, in
section II, we derive the formulation for the mixed states where
we only concentrate on the cases of two-state mixing. In Section
III, we present our numerical results for two-state mixing, and in
Section IV, we discuss possible consequences of three-state mixing
and use several figures to illustrate the changes  of the spectra
from the two-state mixing case. And finally in section V, we
discuss some implications and draw our conclusions.

\section{Pentaquark and Molecular State Mixing}

\subsection{Effective Potential of Molecular State}
We postulate that the molecular state only contains two
constituents. The concerned molecular states can be categorized
into $V$-$N$ and $P$-$N$ systems where $V$ and $P$ correspond to a
vector and a pseudoscalar meon, respectively. Thus, the molecular
state can be $KN$ or $K^{*}N$ for  $\Theta^{+}$, $DN$ or $D^{*}N$
for  $\Theta_{c}$, and $BN$ or $B^{*}N$ for $\Theta_{b}$. The more
complicated structures with three or more constituents will be
commented on later.

We use the traditional method \cite{landaue} by assuming the
potential between a nucleon and a meson to be due to one particle
exchange which may be a scalar, a pseudoscalar, or a vector meson,
and neglecting other heavier and multi-particle intermediate
states. In the linear $\sigma$-model, the effective Lagrangian
relevant to a $\sigma$, a $\pi$ and a $\rho$ exchange is given
by\cite{Geogi,lin,Gokalp,Pir,Bramon,Barry}
\begin{eqnarray}
L&=&g\bar{\psi}(\sigma+i\gamma_5{\mbox{\boldmath $
\tau$}}\cdot{\mbox{\boldmath $\pi$}})\psi
+g_{_{NN}\rho}\bar\psi\gamma_{\mu}{\mbox{\boldmath $ \tau$}}\psi
\cdot{\mbox{\boldmath $ \rho$}}^{\mu} +
g_{PP\sigma}P^{\dagger}P\mathbf{\sigma}+
g_{PP\rho}(P^{\dagger}{\mbox{\boldmath $
\tau$}}\partial_{\mu}P-\partial_{\mu}P^{\dagger}{\mbox{\boldmath $
\tau$}} P)\cdot{\mbox{\boldmath $
\rho$}}^{\mu}\nonumber\\
&&+g_{VV\pi}\varepsilon^{\mu\nu\alpha\beta}\partial_{\mu}V^{\dagger}_{\nu}
{\mbox{\boldmath $
\tau$}}\partial_{\alpha}V_{\beta}\cdot{\mbox{\boldmath $\pi$}}
+g_{VV\sigma}[\partial^{\mu}{V^{\dagger}}^{\nu}\partial_{\mu}
V_{\nu}-
\partial^{\mu}{V^{\dagger}}^{\nu}\partial_{\nu} V_{\mu}]\sigma+g_{VV\rho}
[(\partial_{\mu}
{V^{\dagger}}^{\nu}{\mbox{\boldmath $
\tau$}}V_{\nu}\nonumber\\&&-{V^{\dagger}}^{\nu}{\mbox{\boldmath $
\tau$}}\partial_{\mu}V_{\nu}) \cdot{\mbox{\boldmath $
\rho$}}^{\mu}+({V^{\dagger}}^{\nu}{\mbox{\boldmath $
\tau$}}\cdot\partial_{\mu}{\mbox{\boldmath $ \rho$}}_{\nu}
-\partial_{\mu}{V^{\dagger}}^{\nu}{\mbox{\boldmath $
\tau$}}\cdot{\mbox{\boldmath $ \rho$}}_{\nu})V^{\mu}+
{V^{\dagger}}^{\mu}({\mbox{\boldmath $
\tau$}}\cdot{\mbox{\boldmath $ \rho$}}^{\nu}\partial_{\mu}V_{\nu}-
{\mbox{\boldmath $ \tau$}}\cdot\partial_{\mu}{\mbox{\boldmath $
\rho$}}^{\nu}V_{\nu})]\nonumber\\&&
+[g_{VP\pi}{V^{\dagger}}^{\mu}{\mbox{\boldmath $ \tau$}}
\cdot(P\partial_{\mu}{\mbox{\boldmath $
\pi$}}-\partial_{\mu}P{\mbox{\boldmath $ \pi$}})+h.c.]+g_{VP\rho}
\varepsilon^{\mu\nu\alpha\beta}[\partial_{\mu}{\mbox{\boldmath $
\rho$}}_{\nu}\partial_{\alpha}V^{\dagger}_{\beta}\cdot{\mbox{\boldmath
$ \tau$}}P+\partial_{\mu}V^{\dagger}_{\nu}{\mbox{\boldmath
$\tau$}}\cdot\partial_{\alpha}{\mbox{\boldmath
$\rho$}}_{\beta}P],\nonumber\\
\end{eqnarray}
where $P,V$ are an iso-spin doublet psudoscalar and vector mesons,
i.e. $((K^+)^{(*)},(K^0)^{(*)})^T$, $((D^+)^{(*)},(D^0)^{(*)})^T$,
 $((B^+)^{(*)},(B^0)^{(*)})^T$
and their charge-conjugates. In this expression, we only keep the
concerned terms of the chiral Lagrangian for later calculations.

In analog to the treatment with the chiral Lagrangian, in this
work all the coefficients at the effective vertices are derived by
fitting data of certain physical processes, where all external
particles are supposed to be on their mass-shells. Meanwhile, we
introduce form factors to compensate the off-shell effects of the
exchanged meson. At each vertex, the form factor is parameterized
as \cite{form factor}
\begin{eqnarray}
\frac{\Lambda^{2}-M_{m}^{2}}{\Lambda^{2}-q^{2}}
\end{eqnarray}
where $\Lambda$ is a phenomenological parameter. If the exchanged
particle is on-shell $q^2=M^2_{m}$, the form factor is unity.

To derive an effective potential, we set $q_{0}=0$ and write down
the elastic scattering amplitude in the momentum space and then
carry out a Fourier transformation turning the amplitude into an
effective potential in the configuration space. The total
effective potential for P-N system is the sum of contributions of
$\sigma$ and $\rho$:
\begin{eqnarray}
V_{eff}^{\mathbf{P}-N}(r)=V_{\sigma}^{\mathbf{P}-N}(r)+V_{\rho}^{\mathbf{P}-N}(r),\label{v1}
\end{eqnarray}
where $V_{\sigma}^{\mathbf{P}-N}(r)$ and
$V_{\rho}^{\mathbf{P}-N}(r)$ are the parts of the potential
induced by exchanging $\sigma$ and $\rho$ mesons respectively.

For a $V$-$N$ system, the effective potential is obtained by
exchanging $\pi$, $\sigma$ and $\rho$ mesons. Thus the total
effective potential is the sum of these contributions,
\begin{eqnarray}
V_{eff}^{\mathbf{V}-N}(r)=V_{\pi}^{\mathbf{V}-N}(r)+V_{\sigma}^{\mathbf{V}-N}(r)
+V_{\rho}^{\mathbf{V}-N}(r).\label{v2}
\end{eqnarray}
The  explicit expressions of the individual potentials
$V_{\sigma,\pi, \rho}^{\mathbf{V}-N}(r)$ are given in the
Appendices A.

Using the above potential and the Schr\"odinger equation
\begin{equation}\label{sch}
\Bigg[\frac{\mathbf{p}^2}{2\mu}+V(r)\Bigg]\Psi_{p}(r)=E_{p}\Psi_{p}(r),
\label{scr}
\end{equation}
one can obtain the binding energies of the molecular states. We
suppose the parity of $\Theta^+$ (as well as $\Theta_c$ and
$\Theta_b$) to be positive as predicted in
Ref.\cite{Jaffe,Lipkin1,Lipkin2}, therefore $P$-$N$ and $V$-$N$
must reside in the P-states, i.e. $l=1$. In the above $\mu$ is the
reduced mass of the $P$-$N$ or $V$-$N$ systems. The binding
energies $E_{Mole}$ obtained from the above for different systems
and the corresponding masses $M_{Mole}$ of the molecular states
are given in Table \ref{mole}.

\subsection{The Mixing Mechanism}
In this subsection, we only discuss the mixing between the
pentaquark state with one molecular state which can be either P-N
or V-N type.

We see from Table \ref{mole} that none of pure molecular state has the
right mass for an observed $\Theta$ state. We now discuss how
mixing of the pure pentaquark and molecular state can modify the
masses and obtain the correct ones by assuming two state mixing.
With mixing, the Hamiltonian for the two-state quantum system has
the form
\begin{eqnarray}
 H=\left ( \begin{array}{l l}M_{Mole}&\hspace{0.4cm}\Delta \\
 \hspace{0.4cm}\Delta^{*}&M_{Penta}
 \end{array}
 \right ),
 \end{eqnarray}
where $M_{Mole}$ and $M_{Penta}$ are the masses for the pure
pentaquark and molecular states. $\Delta$ is a mixing parameter.
It is related to non-perturbative QCD and not calculable so far
which we treat as a phenomenological parameter to be determined by
fitting data.

\begin{figure}[h]
\begin{center}
\begin{tabular}{ccc}
\scalebox{0.8}{\includegraphics{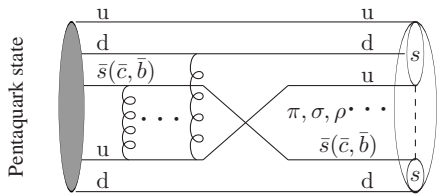}}
%\hspace{1cm}&\hspace{1cm}\scalebox{0.6}
%{\includegraphics{color-exchange.eps}}\hspace{0.5cm}\hspace{1cm}&\hspace{1cm}
%\scalebox{0.6}{\includegraphics{di-tri.eps}}\\
%\\(a)&(b)&
%(c)\\
\end{tabular}
\end{center}
\caption{The diagrams for pentaquark and molecular
state mixing.} \label{co}
\end{figure}

The mixing parameter $\Delta$ is expected to be non-zero. It can
be understood as the following. Suppose we take the
triquark-diquark picture for the pentaquark, the mixing of the
pentaquark and the molecule is due to exchange of an anti-strange
quark in the triquark and a $u$ or $d$ in the diquark accompanied
by gluon exchanges. This mixing effect is related to the
transition process of a pure pentaquark into a nucleon and a
pseudoscalar or a vector meson (if it is kinematically allowed,
the transition can result in a real decay mode) which is depicted
in Fig. \ref{co}.

In Fig. \ref{co}, one can observe that the pentaquark and molecular state
have different color structures. For different models
\cite{Jaffe,Lipkin1,Lipkin2}, the pentaquark may be of the
diqaurk-diquark-anti-strange-quark (or $\bar c,\;\bar b$) and
triquark-diquark sub-structures, whereas the molecular state is
composed of two color-singlet constituents. The mechanism for the
mixing of pentaquark and molecular state is realized via
exchanging multi-gluons and a color re-combination process.
Indeed, for various models\cite{Jaffe,Lipkin1,Lipkin2}, the color
factors would be a bit different.

Diagonalizing $H$, we obtain two real eigenvalues
\begin{eqnarray*}
M_{\pm}=\frac{M_{Mole}+M_{Penta}\pm
\sqrt{(M_{Mole}-M_{Penta})^2+4|\Delta|^2}}{2}.
\end{eqnarray*}
where $M_+$ and $M_-$ correspond to the ``+'' and ``$-$'' on the
right of the above equation. It is noted that $M_-$ is smaller
than $Min(M_{Mole},M_{Penta})$ and $M_+$ is larger than
$Max(M_{Mole},M_{Penta})$. Thus, we can expect that although the
pure pentaquark and molecular states do not have the correct mass,
the mixed state, which corresponds to the observed resonance, can
possess a mass which is consistent with data. If both pure
pentaquark and molecular states are below the observed mass, one
should identify $M_+$ to be the observed one. If both masses are
larger than the observed one, one must identify $M_-$ to be the
observed one. It is not possible to obtain the correct mass if the
observed one is between the pure pentaquark and molecular state
masses.

The eigenstates $|\Psi_+\rangle$ and $|\Psi_-\rangle$
corresponding to $M_+$ and $M_-$, respectively, are written as
\begin{eqnarray*}
|\Psi_{+}\rangle&=&\cos\theta |\mbox{Mole}\rangle
+ \sin\theta e^{i\delta_\Delta}|\mbox{Penta}\rangle,\\
|\Psi_{-}\rangle&=&-\sin\theta |\mbox{Mole}\rangle+ \cos\theta
e^{i\delta_\Delta}|\mbox{Penta}\rangle, \label{mixngstate}
\end{eqnarray*}
where
\begin{eqnarray*}
\cot\theta={|\Delta|\over M_+-M_{Mole}},\;\;e^{i\delta_\Delta} =
{\Delta \over |\Delta|}.
\end{eqnarray*}
The absolute value of $\Delta$ is determined by fitting the
observed state $\Theta$ mass, but the phase $\delta_{\Delta}$ of
$\Delta$ cannot be determined this way.

\subsection{ The Width of the Mixed State}
In the above we have obtained the masses $M_{\pm}$ of the mixed
states, one of which should be consistent with the mass of the
observed resonance and should also produce the observed width. Now
let us turn to the evaluation of the width for the resulting
eigenstate.

For the decay of a mixed state transiting into a two-particle
final state, the rate is given by
\begin{eqnarray}
\Gamma_{\pm}=\int\frac{d^3P_{B}}{(2\pi)^{3}2E_{B}}
\frac{d^3P_{C}}{(2\pi)^{3}}\frac{m_C}{E_{C}}
(2\pi)^4\delta^{4}(P_{A}-P_{B}-P_{C})|\mathcal{A}_{\pm}|^2.\label{gamma}
\end{eqnarray}
where $P_A,\;P_B,\; P_C$ are the four-momenta of the mixed state
and two final products and the amplitude is
\begin{eqnarray}
\mathcal{A}_i=\langle B,C|H_{I}|\Psi_{i}\rangle={\mathcal
D}_{i}\langle B,C|H_{I}^{M}|\mbox{Mole}\rangle+{\mathcal F}_{i}
\langle B,C|H_{I}^{P}|\mbox{Penta}\rangle,\;\;i=+,-,\label{M}
\end{eqnarray}
and
$\mathcal{D}_{+}=\mathcal{F}_{-}e^{-i\delta_\Delta}=\cos\theta$,
$-\mathcal{D}_{-}=\mathcal{F}_{+}e^{-i\delta_\Delta}=\sin\theta$.
$H_{I}^{M}$ acts on the molecular state whereas $H_{I}^{P}$ acts
on the pentaquark state only. They are the interaction Hamiltonian
causing a molecular state and a pure pentaquark state decay to
$B + C$.

We cannot theoretically calculate $\langle
B,C|H_I^{P}|\mbox{Penta} \rangle$ because of its complicated
structure and non-perturbative QCD behavior, but by analyzing its
general property, we relate $\langle
B,C|H_I^{P}|\mbox{Penta}\rangle$ to $\langle
B,C|H_I^{M}|\mbox{Mole}\rangle$, by accounting for their color
structures and physical differences. Thus we may associate the two
amplitudes and write their ratio as
\begin{eqnarray}
\frac{\langle B,C|H_I^{P}|\mbox{Penta}\rangle}{\langle
B,C|H_I^{M}|\mbox{Mole}\rangle}=g\beta,
\end{eqnarray}
where $\beta$ is a corresponding color factor which can be
obtained from Fig.\ref{co}. by considering the color wave function
overlaps. We find that $|\beta| = 2/3\sqrt{3}$ for the
diquark-diquark-antiquark model, and $5\sqrt{2}/3\sqrt{3}$
 for the triquark-diquark model where the leading
contribution is from one gluon exchange. The difference of the
amplitudes $\mathcal{A}(\mbox{Mole} \rightarrow B,C)$ and
$\mathcal{A}(\mbox{Penta} \rightarrow B,C)$ is not only due to the
color factors, but also there may exist a dynamic factor $g$
induced by the concrete physical mechanisms which depend on the
system concerned.  However, they are related to non-perturbative
QCD and cannot be reliably calculated so far, therefore we
introduce an adjustable phenomenological parameter $g$ to denote
the difference of the governing physical mechanisms in the two
transition processes. We will label $g\beta$ by $g_j\beta$ with $j
= s,c,b$ for $\Theta^+$, $\Theta_c$ and $\Theta_b$ seperately.

$\mathcal{A}_{i}$ can be written as
\begin{eqnarray}\label{so}
\mathcal{A}_{i}=\Big(\mathcal{D}_{i}+\mathcal{F}_{i}\cdot
g_{j}\beta \Big) \mathcal{M}(\mbox{Mole}\rightarrow B,C).
\end{eqnarray}
The amplitude $\mathcal{M}(\mbox{Mole}\rightarrow B,C) =
\langle B,C|H_I^{M}|\mbox{Mole}\rangle$ which only
concerns hadronic states, is calculable in terms of the linear
$\sigma$-model, thus with eq.(\ref{so}), one can obtain the
transition amplitude $\mathcal A_i$.

One of the challenging problems with $\Theta^+$ is to explain the
narrowness of the width. There have been many efforts trying to
understand this. If the parameter $g_j$ is of order one, the width
of the pure pentaquark is not necessarily small which seems to
make the situation worse. However when there is mixing, this
problem can be easily solved if the nature selects
$\mathcal{D}_{i} + \mathcal{F}_{i}g_j\beta$ to be small for the
observed state. As a result the other mass eigenstate would have a
broad width.  For later convenience, we define
$$x_j = g_j
\beta exp(i\delta_{\Delta_j}).$$  Using the conjecture that the
physical pentquark state acquires a narrow width by cancellation,
one can determine the combination.

\begin{figure}[h]
\begin{center}
\begin{tabular}{ccc}
\scalebox{0.6}{\includegraphics{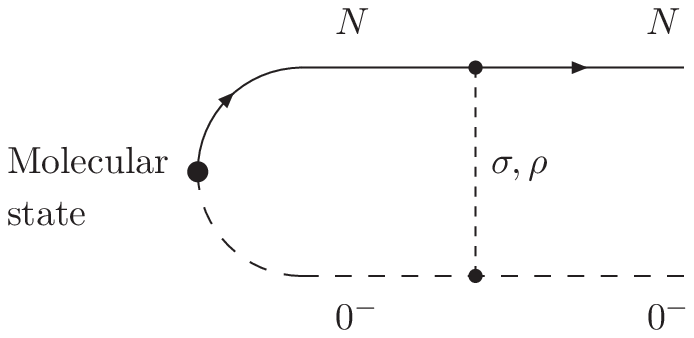}}\hspace{1cm}&\hspace{1cm}
\scalebox{0.6}{\includegraphics{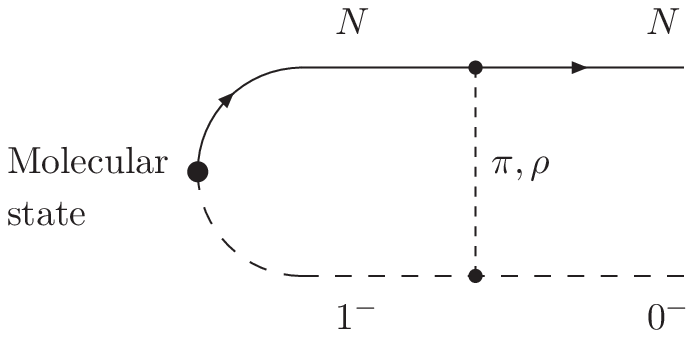}}\hspace{1cm}&\hspace{1cm}
\scalebox{0.6}{\includegraphics{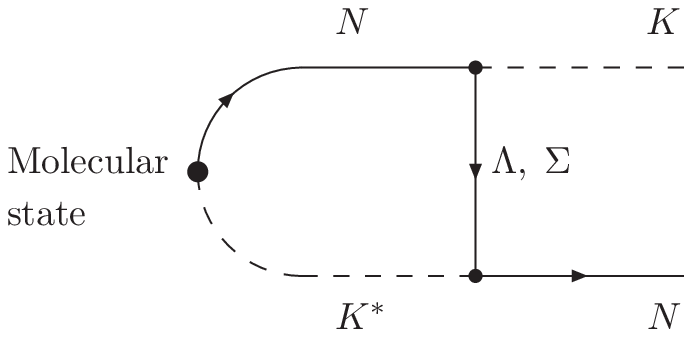}}\\
\hspace{2cm}(a)\hspace{0.3cm}&\hspace{3cm}(b)&\hspace{2cm}(c)
\end{tabular}
\end{center}
\caption{The diagrams for $\Theta^+$ decays. (a) and
(b) correspond to the molecular states which are of $P$-$N$ or
$V$-$N$ types. The pseudoscalar $K$($0^-$) should be replaced by
$D$ or $B$ and $K^*$($1^-$) should be replaced by $D^{*}$ or
$B^{*}$ for $\Theta_c$ and $\Theta_b$
respectively.}\label{transition}
\end{figure}

The molecular decay processes are depicted in
Fig.\ref{transition}. It is noted that the transition of
$|\mbox{Mole}\rangle$ to $N$ and $P$ can also take place via
exchanging a $\Lambda$ or $\Sigma$ baryon, but since they are
heavier than $\pi$, $\sigma$ and $\rho$, the corresponding
contributions are suppressed and we ignore them in our practical
computations.

We will use harmonic oscillator model\cite{harmonic} to estimate
the decay amplitude of a molecular state. The detail expressions
are listed in Appendix B.

\subsection{ The Mass and Width for $\Theta_{b}$}
The above results for mixed state can also be applied to the
$\Theta_b$ state. As indicated above, $\Delta_s$ may be completely
different from $\Delta_c$. For the same reason $g_{ s}$ is
expected to be different than $g_c$, but one can expect $\Delta_c$
and $g_{c}$ are related to $\Delta_b$ and $g_{b}$ since both
$\Delta$ and $g$ are due to an exchange of a heavy anti-quark
($\bar c$ or $\bar b$) with a quark accompanied by a gluon
exchange for the leading order.
By the one-gluon-exchange (OGE) mechanism \cite{OGE},
one may expect that the leading order of the effective potential
is approximately proportional to the distance between the two
constituents (here they refer to a light quark in the diqaurk and
a heavy quark in the triquark) and thus inversely proportional to
the reduced mass. We can roughly have
\begin{eqnarray}
\frac{\Delta_{b}}{\Delta_{c}}&=&\frac{m_{\mu}(BN)}{m_{\mu}(DN)}
=\frac{m_{B}(m_{D}+m_{N})}{m_{D
}(m_{B}+m_{N})}=1.28,\nonumber\\
\frac{\Delta_{b}}{\Delta_{c}}&=&\frac{m_{\mu}(B^{*}N)}
{m_{\mu}(D^{*}N)}=\frac{m_{B^*}(m_{D^*}+m_{N})}{m_{D^*
}(m_{B^*}+m_{N})}=1.17,\label{re}
\end{eqnarray}
here $m_{\mu}(DN)$, $m_{\mu}(BN)$, $m_{\mu}(D^* N)$ and
$m_{\mu}(B^{*}N)$ are respectively the reduced masses of $DN$,
$BN$, $D^{*}N$ and $B^* N$ system. We use similar relation for
$g_b/g_c$.

\section{Numerical Result}
We are now ready to carry out numerical analysis. For the on-shell
vertex parameters involved, we follow
references\cite{Lin,Wujun,Bracco,Deandrea,Casal,9901431,0208168}
to use:

$g_{NN\pi}=g_{NN\sigma}=13.5$, $g_{NN\rho}=3.25$ \cite{Lin}.

$g_{KK\sigma}=4.50$ GeV; $g_{DD\sigma}=12.0$ GeV;
$g_{BB\sigma}=35.0$ GeV \cite{Wujun}.

$g_{KK\rho}=8.49$ \cite{Bracco,Deandrea}.

$g_{K^*K^*\pi}=g_{K^*K^*\sigma}=8.0$,
$g_{D^*D^*\pi}=g_{D^*D^*\sigma}=3.5$,
$g_{B^*B^*\pi}=g_{B^*B^*\sigma}=4.8$ \cite{Deandrea,Casal},

$g_{D^*D^*\rho}=g_{B^*B^*\rho}=2.9$ \cite{Bracco,Deandrea},

$g_{K^*K^*\rho}=4.8$. $g_{D^{*}D\pi}=18$, $g_{B^{*}B\pi}=49.1$
\cite{9901431}.

$g_{DD\rho}=3.81$, $g_{BB\rho}=5.37$, $g_{D^{*}D\rho}=4.71$
GeV$^{-1}$, $g_{B^{*}B\rho}=5.70$ GeV$^{-1}$ \cite{0208168}.

It is generally believed that the parameter $\Lambda$ in the form
factor is around 1 GeV, but the concrete number can vary in a
certain range. If the value of $\Lambda$ is too small, the two
constituents ($PN$ or $VN$) cannot be bound at all, i.e. the
supposed molecular state does not exist, whereas, if the value of
$\Lambda$ is too large, the binding energy becomes negative. We
will allow $\Lambda$ to vary up to a few GeV. By solving the
Schr\"odinger equation, we notice that for the $PN-$system ($KN,\;
DN,\; BN$), the value of $\Lambda$ can be $1.5\sim 2.5$ GeV, and
for the $VN-$system ($K^* N$, $D^{*}N$, $B^* N$) it is
0.5$\sim$1.5 GeV.

%According to the ranges, we present the
%corresponding binding energies of the $PN$ and $VN$ molecular
%states in Table \ref{mole}.

Solving the Schr\"odinger equation with the potential derived in
the linear $\sigma$-model, we obtain the binding energies for pure
molecular states of $KN$, $K^{*}N$, $DN$, $D^{*}N$, $BN$ and
$B^*N$. The predicted values are listed in Table \ref{mole}. The
masses of the molecular states are given by $M_{Mole} = m_N
+M_{P(V)} + E_{P(V)N}$ which are also listed in Table \ref{mole}.

\begin{table}[htb]
\begin{center}
\begin{tabular}{|c|c|c|c|c|c|} \hline
\multicolumn{3}{|c|}{$P$-$N$ System
}\vline&\multicolumn{3}{|c|}{$V$-$N$ System}
\\\hline\hline
\multicolumn{3}{|c|}{$\Lambda: 1.5\sim 2.5$
GeV}\vline&\multicolumn{3}{|c|}{$\Lambda: 0.5\sim 1.5$ GeV}
\\\hline $E_{KN}$(MeV)&$E_{DN}$
(MeV)&$E_{BN}$ (MeV)&$E_{K^{*}N}$ (MeV)&$E_{D^{*}N}$
(MeV)&$E_{B^{*}N}$ (MeV)\\
\hline
$0\sim 25$&$0\sim 13$&$0\sim 9$&$0\sim 10$&$0\sim 4$&$0\sim 2$\\
\hline $M_{KN}$(MeV)&$M_{DN}$ (MeV)&$M_{BN}$ (MeV)&$M_{K^{*}N}$
(MeV)&$M_{D^{*}N}$
(MeV)&$M_{B^{*}N}$ (MeV)\\
\hline $1432\sim 1457$&
$2803\sim2816$&$6217\sim6226$&$1830\sim1840$&$2945\sim2949$&$6263\sim6265$
\\
\hline
\end{tabular}
\end{center}
\caption{The binding energies $E_{Mole}$ and $M_{Mole}$ for
various systems.}\label{mole}
\end{table}

\subsection{The Mixing Parameters}
Several groups have evaluated the masses of pure pentaquarks in
different models. In our numerical evaluations, for concreteness
we adopt the triquark-diquark structure proposed by Ref.
\cite{Lipkin1,Lipkin2}.
\\

{\bf a. The Results for $\Theta^+$}

The value 1592 MeV for the mass of $\Theta^+$ obtained by Karliner
and Lipkin is greater than the measured value (1540 MeV). To
obtain a lower eigenmass, one must mix it with a state which also
has a mass larger than the observed one. If a state with a lower
mass is used, the resulting lower eigenstate would have a mass
even lower in contradiction with data. This forbids $KN$ molecular
state to be the one to mix with. The state which the pure
pentaquark will mix with should be a molecular state of $K^* N$
type. One should identify $|\Psi_-\rangle$ as the $\Theta^+$
state. By fitting data, we have obtained the mixing parameter
$\Delta_{s}$ and $x_{s}$ and other quantities. We have
\begin{eqnarray}
&&x_{s}=0.46\sim0.57,\;\;\;\;\Delta_s=101\sim 137 \mbox{(MeV)},\nonumber\\
&&\Gamma_{\Theta^+}=0.66\sim 1.26
 \mbox{(MeV)},\;\;\;\;\sin\theta = 0.36\sim 0.41,\nonumber\\
&&M_{+}=1879\sim 1889 \mbox{(MeV)},\;\;\;\; \Gamma_{+}= 103\sim
155 \mbox{(MeV)}.
\end{eqnarray}
Here $M_+$ and $\Gamma_+$ are the mass and decay width of the
partner state of $\Theta^+$ which corresponds to the larger
eigenvalue.

One notes that a state of mass around 1885 MeV and broad width
around 130 MeV is predicted. This state is above the $N$-$K$ and
$N$-$K^*$ threshold and therefore may decay into them by strong
interaction. One immediate question arises, why this state has not
been discovered. There are several factors which may have
contributed to the non-observation of this state if it exist, one
of them is that a messy hadron spectra in that energy region where
the expected resonance is hard to be clearly pinned down and
mis-identified as
background. Of course, at present, we cannot confirm the picture
of mixing, namely it could be wrong and the resonance would be
completely interpreted as a pure pentaquark.

{\bf b. The Results for $\Theta_c$ and $\Theta_b$}

In the case of $\Theta_c$, both molecular masses of $N$-$D$ and
$N$-$D^*$, and also the mass of the pure pentaquark are below the
observed mass, a mixing of the pure pentaquark and molecular
states can give correct mass. We also assume $\Theta_b$ to be in a
similar situation.

(i) The case of $P$-$N$ molecular states

First we suppose that the molecular states of $DN$ and $BN$ mix
with the pure pentaquarks $uudd\bar{c}$ and $uudd\bar{b}$ to
construct $\Theta_c$ and $\Theta_b$. We have
\begin{eqnarray}
\mbox{For the $\Theta_c$ State}:
&&x_{c}=-0.69\sim-0.98,\;\;\;\;\Delta_c=127\sim 229 \mbox{(MeV)},\nonumber\\
&&\Gamma_{\Theta_{c}} = 6.9\sim
12.5\mbox{(MeV)},\;\;\;\;\sin\theta= 0.90\sim 0.80,\nonumber\\
&&M_-=2631\sim 2748\mbox{(MeV)}.\nonumber\\
\mbox{For the $\Theta_b$ State}:
&&x_{b}=-0.69\sim-0.98,\;\;\;\;\Delta_b=162\sim
292\mbox{(MeV)},\;\;\;\;\sin\theta = 0.89\sim 0.78\nonumber\\
&&M_{\Theta_b}=6458\sim 6647\mbox{(MeV)},\;\;\;\;
\Gamma_{\Theta_{b}}=2.6\sim 1.9\mbox{(MeV)}\nonumber\\
&&M_{-}=5984\sim 6134\mbox{(MeV)}.
\end{eqnarray}

For the above two cases, the larger one of the two eigenvalues
corresponds is the observed $\Theta_c$. $M_-$ is the mass of
another eigenstate which is below the $N$-$D$ and $N$-$B$
threshold and therefore do not have strong decay channels. They
can easily escape the detection. For the charged $\Theta_b$, there
might be a trace of energy deposit on its path in a drift chamber
and this signal may be used to identify its existence.

(ii) The case of $V$-$N$ molecular states

If the molecular states in $\Theta_c$ and $\Theta_b$ are $D^{*}N$
and $B^{*}N$, the results are different from the $P$-$N$ case. We
have
\begin{eqnarray}
\mbox{For the $\Theta_c$
State}:&&x_{c}=-0.81\sim-1.18,\;\;\;\;\Delta_c= 90\sim
139\mbox{(MeV)},
\nonumber\\
&&\Gamma_{\Theta_{c}}=3.3\sim
15.3\mbox{(MeV)},\;\;\;\;\sin\theta =0.85\sim 0.76\nonumber\\
&&M_{-}= 2825\sim 2892\mbox{(MeV)},\;\;\;\Gamma_{-}=53.5\sim109.9\mbox{(MeV)}.\nonumber\\
 \mbox{For the $\Theta_b$ State}:
&&x_{b}=-0.81\sim-1.18,\;\;\;\;\Delta_b=112\sim
173\mbox{(MeV)},\;\;\;\;\sin\theta=0.91\sim 0.79,\nonumber\\
&&M_{\Theta_b}=6426\sim 6552\mbox{(MeV)},\;\;\;\;
\Gamma_{\Theta_b}=3.1\sim 12.7\mbox{(MeV)},\nonumber\\
&&M_-= 6128\sim 6211\mbox{(MeV)}.
\end{eqnarray}

Again the larger one of the two eigenvalues corresponds to the
observed $\Theta_{c,b}$ and $M_-$ is the mass of another
eigenstate. It is interesting to note that in this case the light
partner of $\Theta_b$ is below the threshold of $N$-$B$ and
therefore has no strong decay channel, but the light partner of
$\Theta_c$ is above the $N$-$D$ threshold and can decay into $N +
D$ by strong interaction via the diagram shown in Fig. \ref{transition}(c).
This
state however has a broad width which may be difficult to
identify. If future experiments with high precision still do not
discover such a state, the mixing of $D^*$-$N$ molecular state
with a pure pentaquark should be ruled out.

\section{Multi-state mixing}

As pointed out in the introduction, there could be multi-state
mixing among pentaquark and molecular states of N-P and N-V types.
For example the mechanisms shown in Fig.2 (b) and (c) can also mix
the N-P and N-V states. By adjusting relevant parameters, the
measured values can be easily re-produced. Allowing pentaquark,
N-P and N-V states to mix, the effective Hamiltonian can be
parameterized as
\begin{eqnarray}
H= \left ( \begin{array}{ccc}
M_{Penta}&\Delta_1& \Delta_2\\
\Delta_1^*&M_{PN}&\epsilon\\
\Delta_2^*&\epsilon^*&M_{VN}
\end{array}\label{three}
\right )\;,
\end{eqnarray}
where $\Delta_{1,2}$, and $\epsilon$ are the parameters describing
the mixing among pentquark and molecular states. Now, the
hamiltonian is expressed by a $3\times 3$ matrix instead of
$2\times 2$ matrix discussed in last section.

In the cases discussed in the previous section the pure pentaquark
and molecular states have masses significantly different from that
of the observed states. This implies that the mixing needed to
explain the data is large. The mixing depends on the size of the
parameter $\Delta_i$ which is of order 100 MeV. The parameter
$\epsilon$ which mixes the N-P and N-V states can be obtained in
our approach by calculating diagrams Fig.2 (b) and (c). We find
that the parameter $\epsilon$ is of order a few MeV which is
considerably smaller than $\Delta_i$ needed to explain data.
Neglecting the mixing between N-P and V-P in our analysis, i.e.
setting $\epsilon$ to be zero, will not affect the main features
of the results.  We will take this simple case to illustrate how
we can obtain the correct masses and correlations of the mixing
parameters with the three-state mixing.

With the above hamiltonian, there are three eigenstates with one
of them being identified as the observed physical states
($\Theta^+,\;\Theta_c$ and possible $\Theta_b$). There may exist
two other physical states. These states have not been discovered
may be due to the same reasons discussed earlier for the other
physical state in the case of two state mixing.

(a) For $\Theta^+$.

The observed state ($\Theta^{+}$) must be identified as the state
with the middle eigenmass $M_M$ shown in Fig.3 (a). To fulfill
this requirement, we must restrict the two mixing parameters
$\Delta_1^{(s)}$ and $\Delta_2^{(s)}$ within a certain range.
Fig.3 (b) demonstrates the relation between $\Delta_1^{(s)}$ and
$\Delta_2^{(s)}$ and ranges for them. We use $M_{H}$ and $M_{L}$
to denote the masses corresponding to the heavier and the lighter
physical states(see Fig. 3). The bands described in Fig. 3 (a) and
(b) come from the experimental error of $M_{\Theta^{+}}$ and the
theoretical uncertainties of the binding energies of $KN$ and
$K^{*}N$ systems. One notes that the P-N state can also play
significant role in the mixing.

\begin{figure}[htb]
\begin{center}
\begin{tabular}{cc}
\scalebox{0.7}{\includegraphics{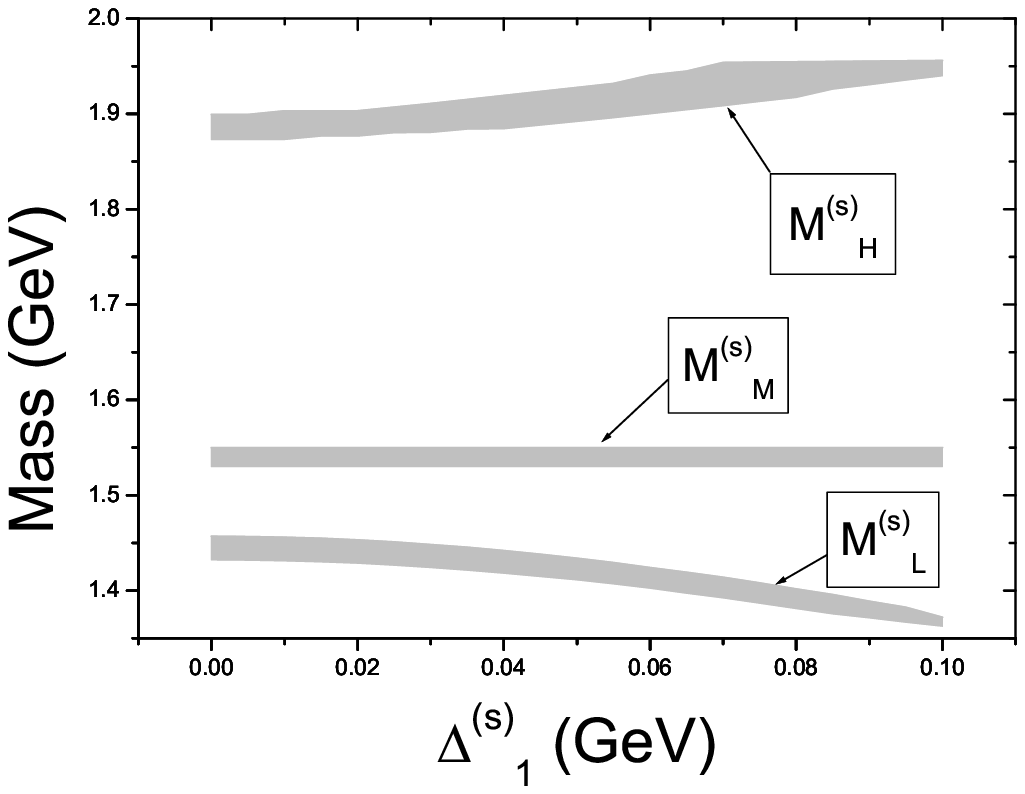}}&\scalebox{0.7}{\includegraphics{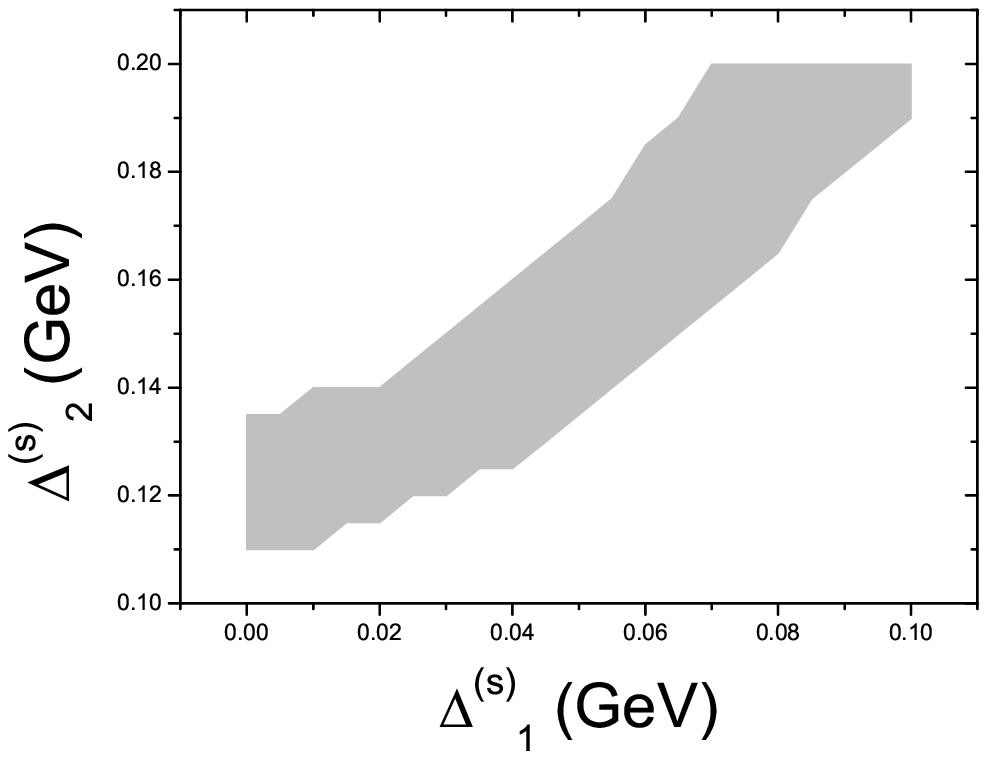}}\\
(a)&(b)
\end{tabular}
\end{center}
\caption{(a) and (b) describe the dependence of $M_{L}^{(s)}$,
$M_M^{(s)}$ and $M_{H}^{(s)}$ on $\Delta_1^{(s)}$, and relation
between $\Delta_{1}^{(s)}$ and $\Delta_{2}^{(s)}$, respectively.
The state with middle eigenmass corresponds to the observed state
$\Theta^{+}$.}
\end{figure}

(b) For $\Theta_{c}$.

Different from the case of $\Theta^{+}$,  the largest one among
the three physical states  corresponds to the observed
$\Theta_{c}$, when we diagonalize the three-states mixing
hamiltonian (\ref{three}). $M_{M}^{(c)}$ and $M_{L}^{(c)}$ are
other two physical states having  middle and lower eigenmasses
respectively. Similarly, we also use two diagrams to demonstrate
the relations between $M_{i}^{(c)}$ and $\Delta_1^{(c)}$, and
$\Delta_{1}^{(c)}$ and $\Delta_{2}^{(c)}$ (see Fig. 4 (a) and
(b)).
\begin{figure}[htb]
\begin{center}
\begin{tabular}{cc}
\scalebox{0.7}{\includegraphics{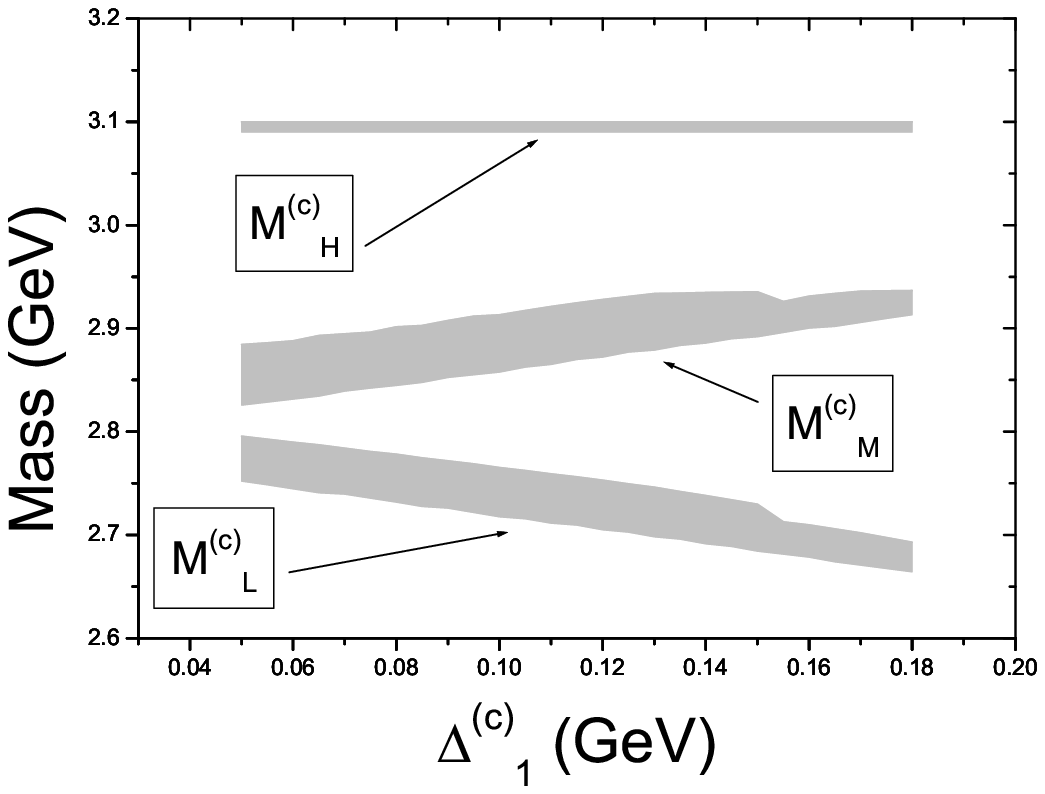}}&\scalebox{0.7}{\includegraphics{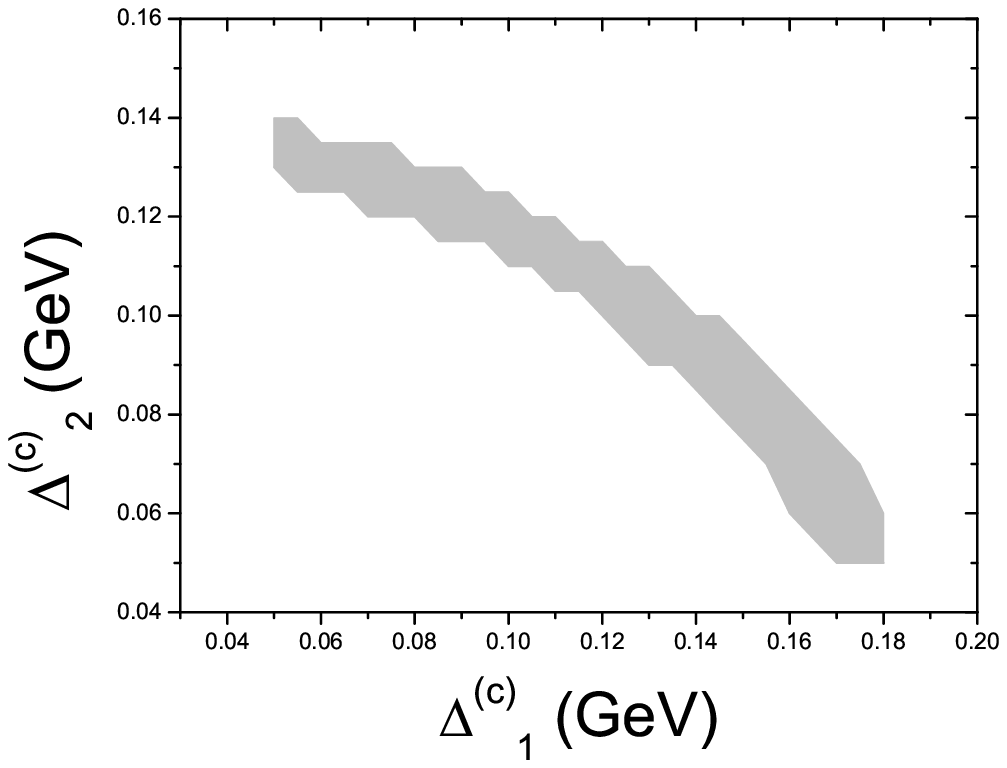}}\\
(a)&(b)
\end{tabular}
\end{center}
\caption{(a) and (b) describe  the dependence of $M_{L}^{(c)}$,
$M_{M}^{(c)}$ and $M_H^{(c)}$ on $\Delta_1^{(c)}$, and the
relation between $\Delta_{1}^{(c)}$ and $\Delta_{2}^{(c)}$,
respectively. The highest state corresponds to the observed
$\Theta^{c}$.}
\end{figure}

(c) For $\Theta_{b}$.

In analog to the case of two-state mixing in last section, we
apply the relations (\ref{re}) of $\Delta_{1}^{(c)}$,
$\Delta_{2}^{(c)}$ with $\Delta_{1}^{(b)}$, $\Delta_{2}^{(b)}$ to
predict three physical states, whose masses are denoted as
$M_{H}^{(b)}$, $M_{M}^{(b)}$ and $M_{L}^{(b)}$ respectively. We
may expect that the state having the largest eigenmass
$M_{H}^{(b)}$ corresponds to $\Theta_{b}$ which is a counterpart
of the observed $\Theta_c$. In Fig. 5, we draw a diagram depicting
the relations of the masses of the three physical states.
\begin{figure}[htb]
\begin{center}
\scalebox{0.7}{\includegraphics{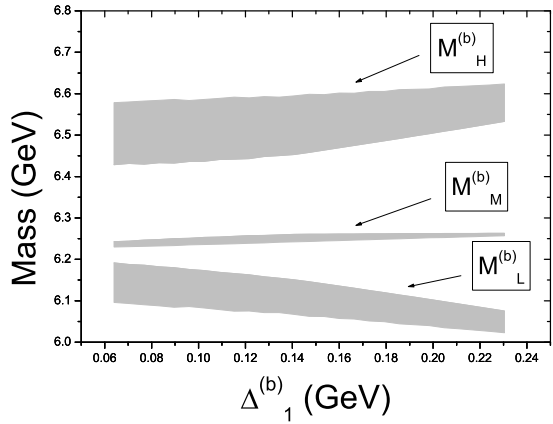}}
\end{center}
\caption{The dependence of $M_{L}^{(b)}$, $M_{M}^{(b)}$ and
$M_H^{(b)}$ on $\Delta_1^{(b)}$ }
\end{figure}

Whether there is significant three state mixing, that is both
$\Delta_{1}$ and $\Delta_2$ are sizeable, has to be determined by
future experimental data.

\section{Conclusion and discussion}

In this work, motivated by the fact that the theoretically
evaluated masses and widths of pure pentaquark or molecular states
do not coincide with the observed $\Theta^+$ and $\Theta_c$
states, we have studied some consequences by assuming that the
observed resonances $\Theta^+$ and $\Theta_c$ are mixtures of pure
pentaquark and molecular states.

The pure pentaquarks may be in the triquark-diquark or
diquark-diquark-antiquark structures, while the molecular states
in fact are only a re-combination of the quark constituents  and
colors, i.e. another component in the Hilbert space. Therefore a
mixing between the molecular state and pure pentaquark is
possible. Combining theoretical estimates for the masses of the
pure pentaquark given in the literatures and our estimate for the
masses of pure molecular states in the linear $\sigma$-model, we
estimated the mixing parameters, $\Delta_i$ (here i=s,c) by
fitting data.

We find that through the mixing mechanism it is possible to obtain
the observed masses for $\Theta^+$ and $\Theta_c$, and also
possible to obtain narrow widths for these states through
destructive interferences even if the pure pentaquark and the
molecular states may have broader decay widths. The mixings are
sizeable, but the dominant components of the observed states are
pentaquark states.

An interesting prediction of mixing of pure pentaquark and
molecular states is that there exists another physical state. In
the case of $\Theta^+$, with the pure pentaquark mass predicted by
the triquark-diquark model\cite{Lipkin1},
the state to be mixed is the $N$-$K^*$ molecular
state. The resulting heavier physical state mass is predicted to
be in the range $1879\sim 1889$ MeV with a width in the range
$103\sim 155$ MeV. Since this state is above the producction threshold
of $N$-$K$ and $N$-$K^*$, its strong decays into $N$-$K$ and $N$-$K^*$
can be used
to discover such a state. At present there is no evidence for such
a state. It may be due to experimental sensitivity since this is a
region where there is a mess spectra, and this physical state has
a broad width, so that it might be hidden in the forest of hadrons
in the region and is mis-identified as the background. Of course
there is also the possibility that the mixing for $\Theta^+$ is
not needed and the pure pentaquark state has the right properties
as attempted by many investigations.

For $\Theta_c$, it is another story, by contraries. If the pure
pentaquark mixes with a $N$-$D$ molecular state, the mixing
mechanism would predict that the mass of the other state is below
the threshold of $N$-$D$. This state is stable against strong
interaction, and may have escaped detection in the detector. There
may be several weak decay channels, but difficult to detect
either. Whereas if the pure pentaquark mixes with a $N$-$D^*$
state the light partner of $\Theta_c$ is above the $N$-$D$
threshold and can decay into $N + D$ by strong interaction. This
state however has a broad width which may be difficult to
identify. If future experiments with high precision still do not
discover such a state, the mixing of $D^*$-$N$ molecular state
with a pure pentaquark should be ruled out.

For $\Theta_b$ the light partner state is below the $N$-$B$
threshold for both the cases that the pure pentaquark mixes with a
$B$-$N$ or $B^*$-$N$ molecular state. Since the light partner
state is charged, although it does not have strong decay modes, it
may leave trace by depositing energy in the medium when passing
through a detector, such as a drift chamber. We encourage our
experimental colleagues to carry out a search in the relevant
region.

Obviously there could be multi-state mixing among
diqaurk-diqaurk-antiqaurk, diqaurk-triquark and molecular
state(s). By adjusting parameters (there are more of them than in
the two-state mixing), the measured values can be re-produced. In
section IV, we illustrate possible changes if three-state mixing
is considered. We find that for the present experimental data, it
is easy to restore the case for two-state mixing by requiring one
of $\Delta_i$ to be zero. Thus the main feature is clearly given
in the two-state mixing case. Since we cannot reliably evaluate
the mixing parameter from any solid theoretical ground,
considering mixing among more states does not provide us with
further information. At present, the two-state mixing can result
in values which well explain the spectra and narrow widths of
$\Theta^+$, $\Theta_c$ and predict possible $\Theta_b$. However,
in the future more accurate measurements on properties of the
resonances may demand such multi-state mixing.

As a conclusion, a mixing between a pure pentaquark and a
molecular state may be reasonable and by this picture, we can
explain the mass spectra and widths of the observed $\Theta^+$ and
$\Theta_c$ even the theoretical estimations based on the pure
pentaquark given in the literatures  obviously deviate from data.
Applying the same mechanism, we have predicted the mass and width
of $\Theta_b$ which can be tested in the future experiments.
Moreover, multi-state mixing may be required when more accurate
measurements are made in the future.

\vspace{1cm} \noindent Acknowledgment:

We would like to thank professors Lipkin, Toki,  K.T. Chao, B.Q.
Ma, B.S. Zou for many useful discussions. This work is partly
supported by the National Natural
Science Foundation of China.\\

\noindent {\bf Appendix A}\\

(i) The effective potential for the nucleon and pseudoscalar meson
system.

(1) $\sigma$ exchange.
\begin{eqnarray}
  V^{\mathbf{P}-N}_\sigma(\mathbf{q})&=& \frac{-g_{NN\sigma}\,g_{PP\sigma}}
{2m_{a}(\mathbf{q}^2  + {{m_{\sigma }}}^2)}
  \bigg[1 - \frac{{{\mathbf{p}}}_{1}^2}{2\,{m_{a}}^2}-\frac{\mathbf{q}^2}{8m_{N}^2}
  - \frac{\mathbf{q}\cdot\mathbf{p}_{1}+3\mathbf{p}_{1}\cdot\mathbf{q}}
  {8m_{N}^2}- \frac{\mathbf{p}_{1}^2}{2m_{N}^2}
  \nonumber\\&&-\frac{i}{4 m_{N}^2}{\mbox{\boldmath $\sigma$}_2}\cdot
  \left( \mathbf{q}\times\mathbf{p}_{1} \right) \bigg]
  \bigg(\frac{{ {\Lambda }^2 - {{m_{\sigma }}}^2  }}{{ \mathbf{q}^2 + {\Lambda }^2
  }}\bigg)^2,
 \end{eqnarray}

taking the Fourier transformation, we obtain
 \begin{eqnarray}
  V^{\mathbf{P}-N}_\sigma(r)&=&\frac{-g_{NN\sigma}\,g_{PP\sigma}}{2m_{a}}
   \bigg\{f_\sigma(r) -\frac{9}{8m_{N}^2}F_\sigma(r)
   -\bigg( \frac{1}{2\,{m_{a}}^2}
  + \frac{1}{2m_{N}^2}\bigg){{\mathbf{p}}}_{1}^2f_{\sigma}(r)
  \nonumber\\&&+\frac{[\mathbf{\nabla}^2f_\sigma(r)]}{8m_{N}^2}
 +\frac{i\mathbf{p}_{1}\cdot\mathbf{r}}{2m_{N}^2}F_{\sigma}(r)
 -\frac{\mathbf{S}_2\cdot\mathbf{L} }{2 m_{N}^2}F_\sigma(r)
 \bigg\}.
 \end{eqnarray}
where
 \begin{eqnarray*}
  f_\sigma(r)&=&\frac{\displaystyle e^{-m_\sigma r}}{\displaystyle 4\pi r}
  -\frac{\displaystyle e^{-\Lambda r}}{\displaystyle 4\pi r}
  +\frac{\displaystyle (m_\sigma^2-\Lambda^2)e^{-\Lambda r} }{\displaystyle 8\pi\Lambda}, \\
  F_\sigma(r)&=&\frac{1}{\displaystyle r}\frac{\displaystyle \partial}{\displaystyle \partial
  r}f_\sigma(r).
   \end{eqnarray*}

(2) $\rho$ exchange.

 \begin{eqnarray}
 V^{\mathbf{P}-N}_\rho(\mathbf{q})&=&\frac{g_{NN\rho}\,g_{PP\rho}}{\mathbf{q}^2 + {{m_{\rho}}}^2}
  \bigg\{1 - \frac{\mathbf{q}^2}{8m_{N}^2}- \frac{\mathbf{q}\cdot
  \mathbf{p}_{1}-\mathbf{p}_{1}\cdot\mathbf{q}}{8m_{N}^2}
 +\frac{i{\mbox{\boldmath $\sigma$}_2}\cdot(\mathbf{q}\times\mathbf{p}_{1})}{4m_{N}^2}
 +\frac{1}{4m_{N}m_{a}}[\mathbf{q}^2+4\mathbf{p}_{1}^2\nonumber\\&&+2 \mathbf{q}\cdot\mathbf{p}_{1}
 +2 \mathbf{p}_{1}\cdot\mathbf{q}+2i{\mbox{\boldmath $\sigma$}_2}\cdot(\mathbf{q}\times\mathbf{p}_{1})
 ]\bigg\}
  \bigg(\frac {\displaystyle \Lambda^2-m_\rho^2}{\displaystyle \Lambda^2+\mathbf{q}^2}\bigg)^2
\end{eqnarray}
taking the Fourier transformation, we get
 \begin{eqnarray}
  V^{\mathbf{P}-N}_\rho(r)&=&g_{NN\rho}\,g_{PP\rho}
  \bigg\{f_\rho(r)-\frac{3}{8m_{N}^2}F_\rho(r)+\frac{[\mathbf{\nabla}^2f_\rho(r)]}{8m_{N}^2}
  +\frac{\mathbf{S}_2\cdot\mathbf{L}}{2m_{N}^2}F_\rho(r)
  -\frac{[\mathbf{\nabla}^2 f_\rho(r)]}{4m_{N}m_{a}}\nonumber \\&&
+\frac{1}{4m_{N}m_{a}}\left[4 \mathbf{p}_{1}^2
f_\rho(r)+6F_\rho(r) -4i
\mathbf{p}_{1}\cdot\mathbf{r}F_\rho(r)+4\mathbf{S}_2\cdot\mathbf{L}F_\rho(r)
  \right]\bigg\},
  \end{eqnarray}
where
 \begin{eqnarray*}
 f_\rho(r)&=&\frac{\displaystyle e^{-m_\rho r}}{\displaystyle 4\pi r}
  -\frac{\displaystyle e^{-\Lambda r}}{\displaystyle 4\pi r}
  +\frac{\displaystyle (m_\rho^2-\Lambda^2)e^{-\Lambda r} }{\displaystyle 8\pi\Lambda},\\
 F_\rho(r)&=&\frac{1}{\displaystyle r}\frac{\displaystyle \partial}{\displaystyle \partial
 r}f_\rho(r).
 \end{eqnarray*}

(ii) The effective potential for the nucleon and vector meson
system.

(1) pion exchang.

\begin{eqnarray}
  V^{\mathbf{V}-N}_\pi(\mathbf{q})&=& -\frac{g_{NN\pi}\,g_{VV\pi}}{4m_{N}
  (\mathbf{q}^2  + m_{\pi }^2)}
  (\mathbf{S}_1\cdot\mathbf{q})(\mathbf{S}_2\cdot\mathbf{q})
 \bigg(\frac {\displaystyle \Lambda^2-m_\pi^2}{\displaystyle
 \Lambda^2+\mathbf{q}^2}\bigg)^2,
 \end{eqnarray}
taking a Fourier transformation, we get
 \begin{eqnarray}
 V^{\mathbf{V}-N}_\pi(r)&=&\frac{g_{NN\pi}\,g_{VV\pi}}{4m_{N}}
 (\mathbf{S}_1\cdot\mathbf{\nabla})(\mathbf{S}_2\cdot\mathbf{\nabla})
 f_\pi(r),
 \end{eqnarray}
here
 \begin{eqnarray*}
  f_\pi(r)&=&\frac{\displaystyle e^{-m_{\pi} r}}{\displaystyle 4\pi r}
  -\frac{\displaystyle e^{-\Lambda r}}{\displaystyle 4\pi r}
  +\frac{\displaystyle ({m_\pi}^2-\Lambda^2)e^{-\Lambda r} }{\displaystyle
  8\pi\Lambda}.
 \end{eqnarray*}

(2) $\sigma$ exchange.

\begin{eqnarray}
 V^{\mathbf{V}-N}_\sigma(\mathbf{q})&=& -\frac{g_{NN\sigma}\,g_{VV\sigma}\, m_{b}}
  {2(\mathbf{q}^2  + {{m_{\sigma }}}^2)}
  \bigg[-1 + \frac{{{\mathbf{p}}}_{1}^2}{6\,{m_{b}}^2}
  +\frac{2\mathbf{p}_{1}\cdot\mathbf{q}}{3m_{b}^2}
  -\frac{i\mathbf{S}_1\cdot(\mathbf{q}\times\mathbf{p}_{1})}{4m_{b}^2}
  \nonumber\\&&+ \frac{\mathbf{p}_{1}\cdot\mathbf{q}-\mathbf{q}\cdot\mathbf{p}_{1}}{8m_{N}^2}
  + \frac{\mathbf{q}^2}{8m_{N}^2}
  -\frac{i\mathbf{S}_2\cdot(\mathbf{q}\times\mathbf{p}_{1})}{4m_{N}^2} \bigg]
  \bigg(\frac{{ {\Lambda }^2 - {{m_{\sigma }}}^2  }}{{ \mathbf{q}^2 + {\Lambda }^2
  }}\bigg)^2,
 \end{eqnarray}
taking a Fourier transformation, we obtain
 \begin{eqnarray}
 V^{\mathbf{V}-N}_\sigma(r)&=&-\frac{g_{NN\sigma}\,g_{VV\sigma}\, m_{b}}{2}
   \bigg\{-f_\sigma(r)-\frac{3}{8m_{N}^2}F_\sigma(r) + \frac{{{\mathbf{p}}}^2}{6{m_{b}}^2}f_\sigma(r)
   -\frac{\mathbf{S}_1\cdot\mathbf{L}}{4m_{b}^2}F_\sigma(r)
  \nonumber \\&&-\frac{2i\mathbf{p}_{1}\cdot\mathbf{r}}{3m_{b}^2}F_\sigma(r)
   -\frac{[\mathbf{\nabla}^2 f_\sigma(r)]}{8m_{N}^2}
  -\frac{\mathbf{S}_2\cdot\mathbf{L}}{2m_{N}^2}F_\sigma(r)\bigg\},
 \end{eqnarray}
where
 \begin{eqnarray*}
  f_\sigma(r)&=&\frac{\displaystyle e^{-m_\sigma r}}{\displaystyle 4\pi r}
  -\frac{\displaystyle e^{-\Lambda r}}{\displaystyle 4\pi r}
  +\frac{\displaystyle (m_\sigma^2-\Lambda^2)e^{-\Lambda r} }{\displaystyle 8\pi\Lambda}, \\
  F_\sigma(r)&=&\frac{1}{\displaystyle r}\frac{\displaystyle \partial}
  {\displaystyle \partial r}f_\sigma(r).
   \end{eqnarray*}

(3) $\rho$ exchange.

 \begin{eqnarray}
 V^{\mathbf{V}-N}_\rho(\mathbf{q})&=&\frac{g_{NN\rho}\,g_{VV\rho}}{\mathbf{q}^2 + {{m_{\rho}}}^2}
  \bigg[1 - \frac{\mathbf{q}^2}{8m_{N}^2}
  -\frac{\mathbf{p}_{1}\cdot\mathbf{q}-\mathbf{q}\cdot\mathbf{p}_{1}}{8m_{N}^2}
  +\frac{i\mathbf{S}_2\cdot(\mathbf{q}\times\mathbf{p}_{1})}{4m_{N}^2}
  +\frac{\mathbf{q}^2}{4m_{N}m_{1^{-}}}+\frac{\mathbf{p}_{1}^2}{6m_{N}m_{b}}\nonumber\\&&
  +\frac{\mathbf{p}_{1}\cdot\mathbf{q}}{2m_{N}m_{b}} +\frac{\mathbf{q}\cdot\mathbf{p}_{1}}{12m_{N}m_{b}}
  -\frac{i\mathbf{S}_2\cdot(\mathbf{q}\times\mathbf{p}_{1})}{6m_{N}m_{b}}
  -\frac{i\mathbf{S}_1\cdot(\mathbf{q}\times\mathbf{p}_{1})}{8m_{N}m_{b}}\bigg]
  \bigg(\frac {\displaystyle \Lambda^2-m_\rho^2}{\displaystyle
  \Lambda^2+\mathbf{q}^2}\bigg)^2,
\end{eqnarray}
taking a Fourier transformation, we get

 \begin{eqnarray}
  V^{\mathbf{V}-N}_\rho(r)&=&g_{NN\rho}\,g_{VV\rho}
  \bigg\{f_\rho(r)+\frac{3}{8m_{N}^2}F_\rho(r)
 -\frac{1}{2m_{N}m_{b}}F_\rho(r)
   +\frac{[\mathbf{\nabla}^2 f_\rho(r)]}{8m_{N}^2}
  +\frac{\mathbf{S}_2\cdot\mathbf{L}}{4m_{N}^2}F_\rho(r)\nonumber\\&&
  -\frac{[\mathbf{\nabla}^2 f_\rho(r)]}{4m_{N}m_{b}}+\frac{\mathbf{p}_{1}^2}{6m_{N}m_{b}}f_\rho(r)
  -\frac{7i\mathbf{r}\cdot\mathbf{p}_{1}}{12m_{N}m_{b}}F_\rho(r)
  -\frac{\mathbf{S}_2\cdot\mathbf{L}}{6m_{N}m_{b}}F_\rho(r)
  \nonumber\\&&-\frac{\mathbf{S}_1\cdot\mathbf{L}}{8m_{N}m_{b}}F_\rho(r)\bigg\}
  ,
  \end{eqnarray}

here
 \begin{eqnarray*}
 f_\rho(r)&=&\frac{\displaystyle e^{-m_\rho r}}{\displaystyle 4\pi r}
  -\frac{\displaystyle e^{-\Lambda r}}{\displaystyle 4\pi r}
  +\frac{\displaystyle (m_\rho^2-\Lambda^2)e^{-\Lambda r} }{\displaystyle 8\pi\Lambda},\\
 F_\rho(r)&=&\frac{1}{\displaystyle r}\frac{\displaystyle \partial}{\displaystyle \partial
 r}f_\rho(r).
 \end{eqnarray*}

\noindent{\bf Appendix B}\\
The molecular state is expressed as\cite{harmonic}
\begin{eqnarray}
|\phi_{Mole}(\mathbf{P},s)\rangle=A\cdot \sum_{spin}
C^{s}_{s_{1},s_{2}}\chi_{s_{1},s_{2}}\int
d\mathbf{p}_{1}d\mathbf{p}_{2}\psi(\mathbf{p}_{1},\mathbf{p}_{2})\delta^{3}(\mathbf{p}_{1}+
\mathbf{p}_{2})b^{\dag}_{p_{1},s_{1}}a^{\dag}_{p_{2},s_{2}}|0
\rangle,
\end{eqnarray}
where $ C^{s}_{s_{1},s_{2}}$ is the C-G coefficients,
$\chi_{s_{1},s_{2}}$ are the spin-wavefunctions and $A$ is a
normalization constant. We normalize this fermion state as
\begin{eqnarray}
\langle
\phi_{Mole}(\mathbf{P}')|\phi_{Mole}(\mathbf{P})\rangle=(2\pi)^{3}\frac{E_{P}}{M_{A}}\;\delta^{3}
(\mathbf{P}'-\mathbf{P}).
\end{eqnarray}
In Fig.\ref{transition}(a), we present the diagram for decay of
the molecular state which is composed of a pseudoscalar meson and
a nucleon in P-state, this transition occurs via exchanging
$\sigma $ or $\rho$, the amplitudes are
\begin{eqnarray}
\mathcal{M}(\sigma)^{(PN)}&=&A\cdot g_{NN\sigma}g_{PP\sigma}\int
d\mathbf{p}_1
d\mathbf{p}_{2}\sum_{spin}C^{s}_{s_{1},s_{2}}\chi_{s_{1},s_{2}}
\bar{u}(P_{B},s_{B})u(p_{1},s_{1})
\psi(\mathbf{p}_{1},\mathbf{p}_{2})\nonumber\\&&\times\delta^{3}(\mathbf{p}_{1}+
\mathbf{p}_{2})\cdot \frac{1}{q^{2}-m_{\sigma}^2}\bigg(\frac
{\displaystyle \Lambda^2-m_\sigma^2}{\displaystyle
  \Lambda^2-q^2}\bigg)^2,\\
\mathcal{M}(\rho)^{(PN)}&=&A\cdot g_{NN\rho}g_{PP\rho}\int
d\mathbf{p}_1
d\mathbf{p}_{2}\sum_{spin}C^{s}_{s_{1},s_{2}}\chi_{s_{1},s_{2}}
\bar{u}(P_{B},s_{B})\gamma^{\mu}u(p_{1},s_{1})
(P_{C}+p_2)^{\nu}\nonumber\\&&\times
\psi(\mathbf{p}_{1},\mathbf{p}_{2})\delta^{3}(\mathbf{p}_{1}+
\mathbf{p}_{2}) \cdot
\frac{g_{\mu\nu}}{q^{2}-m_{\rho}^2}\bigg(\frac {\displaystyle
\Lambda^2-m_\rho^2}{\displaystyle
  \Lambda^2-q^2}\bigg)^2.
\end{eqnarray}
and the total amplitude is the sum of $\mathcal{M}^{(PN)}(\sigma)$
and $\mathcal{M}^{(PN)}(\rho)$.

In Fig.\ref{transition}(b), the molecular state consists of a
vector meson and a nucleon, the corresponding amplitudes are
\begin{eqnarray}
\mathcal{M}(\pi)^{(VN)}&=&A\cdot g_{NN\pi}g_{VP\pi}\int
d\mathbf{p}_1
d\mathbf{p}_{2}\sum_{spin}C^{s}_{s_{1},s_{2}}\chi_{s_{1},s_{2}}
\bar{u}(P_{B},s_{B})\gamma^{5}u(p_{1},s_{1})
(2P_{C}-p_2)^{\mu}\nonumber\\&&\times
\epsilon_{\mu}\psi(\mathbf{p}_{1},\mathbf{p}_{2})
\delta^{3}(\mathbf{p}_{1}+ \mathbf{p}_{2})\cdot
\frac{1}{q^{2}-m_{\pi}^2}\bigg(\frac {\displaystyle
\Lambda^2-m_\pi^2}{\displaystyle
  \Lambda^2-q^2}\bigg)^2,\\
\mathcal{M}(\rho)^{(VN)}&=&A \cdot g_{NN\rho}g_{VP\rho}\int
d\mathbf{p}_1
d\mathbf{p}_{2}\sum_{spin}C^{s}_{s_{1},s_{2}}\chi_{s_{1},s_{2}}
\bar{u}(P_{B},s_{B})\gamma^{\lambda}u(p_{1},s_{1})
\varepsilon^{\alpha\beta\mu\nu}
q_{\alpha}p_{2\mu}\nonumber\\&&\times\epsilon_{\nu}\psi(\mathbf{p}_{1},\mathbf{p}_{2})
\delta^{3}(\mathbf{p}_{1}+ \mathbf{p}_{2})\cdot
\frac{g_{\lambda\beta}}{q^{2}-m_{\rho}^2}\bigg(\frac
{\displaystyle \Lambda^2-m_\rho^2}{\displaystyle
  \Lambda^2-q^2}\bigg)^2.
\end{eqnarray}
and the total amplitude is the sum of $ \mathcal{M}(\pi)^{(VN)}$
and $\mathcal{M}(\rho)^{(VN)}$.

\end{document}